\documentclass[conference]{IEEEtran}

\IEEEoverridecommandlockouts
\usepackage{amsmath}%
\usepackage{amsfonts}%
\usepackage{amssymb}%
\usepackage{graphicx}
\usepackage{verbatim}%
\usepackage{enumerate}%

\long\def\symbolfootnote[#1]#2{\begingroup%
\def\thefootnote{\fnsymbol{footnote}}\footnote[#1]{#2}\endgroup} 

\begin{document}

\title{Distributed Spectrum Sensing with Sequential Ordered Transmissions to a Cognitive Fusion Center}

\author{\IEEEauthorblockN{Laila Hesham, Ahmed Sultan and Mohammed Nafie}
\IEEEauthorblockA{Wireless Intelligent Networks Center (WINC)\\
Nile University, Cairo, Egypt.\\
E-mail: laila.hesham@nileu.edu.eg, \{asultan, mnafie\}@nileuniversity.edu.eg }}

\maketitle

\begin{abstract}
Cooperative spectrum sensing is a robust strategy that enhances the detection probability of primary licensed users. However, a large number of detectors reporting to a fusion center for a final decision causes significant delay and also presumes the availability of unreasonable communication resources at the disposal of a network searching for spectral opportunities. In this work, we employ the idea of sequential detection to obtain a quick, yet reliable, decision regarding primary activity. Local detectors take measurements, and only a few of them transmit the log likelihood ratios (LLR) to a fusion center in descending order of LLR magnitude. The fusion center runs a sequential test with a maximum imposed on the number of sensors that can report their LLR measurements. We calculate the detection thresholds using two methods. The first achieves the same probability of error as the optimal block detector. In the second, an objective function is constructed and decision thresholds are obtained via backward induction to optimize this function. The objective function is related directly to the primary and secondary throughputs with inbuilt privilege for primary operation. Simulation results demonstrate the enhanced performance of the approaches proposed in this paper. We also investigate the case of fading channels between the local sensors and the fusion center, and the situation in which the sensing cost is negligible.\symbolfootnote[1]{Part of this work was presented in ICASSP, May 2011.} 

\symbolfootnote[0]{This work was supported in part by a grant from the Egyptian NTRA (National Telecommunications Regulatory Authority).}
      
\end{abstract}

\begin{IEEEkeywords}
Cognitive Radios, Cooperative Spectrum Sensing, Detection delay, Sequential detection. 
\end{IEEEkeywords}

\section{Introduction}   Cognitive radio is an emerging technology aimed at solving the problem of spectrum under-utilization caused by static spectrum allocation \cite{haykin}. The technology allows a cognitive, also called unlicensed or secondary, terminal to make efficient use of any available spectrum at any given time. This involves the detection of the activity of licensed, or primary, users so that the secondary operation does not interfere with the primary networks and disrupt their services. This detection task may be extremely difficult due to a large array of factors including, inter alia, the wide variety of primary networks and the uncertainty about the propagation conditions between an active primary terminal and a secondary sensor attempting to detect primary activity. If the channel between a primary transmitter and a secondary sensor is in deep fade, the sensor would falsely decide that the probed channel is vacant. Secondary transmission on this channel would then cause interference on the primary link. Relying on one sensor to detect the presence of primary operation is therefore highly unreliable \cite{SPmag}. 

Enhancement of sensing reliability requires a system of spatially distributed multiple sensors cooperating together in order to mitigate the impact of channel uncertainty provided that the channels between them and the primary transmitter are independent.
Many papers have shown improvement in spectrum sensing capabilities through cooperation between individual
cognitive users (see, e.g., \cite{Akyildiz201140,Ganesan05,ghasemi05,Mishra06,alisayed_optimal_linear}). However, in order to reap the benefits of distributed detection and cooperative diversity, efficient schemes are needed to combine the data obtained by the local sensors \cite{tsit}. There is typically a fusion center at which the final decision is made. Whether the fusion center receives the raw unprocessed data from local sensors or summary messages obtained after some local processing, there are challenges concerning the transfer of data to the fusion center. It may be assumed, for instance, that there are orthogonal channels available for the transmission from the local detectors to the fusion center. In the context of cognitive radio, this assumption might not be practical given that the main objective of the cognitive users is to find communication resources. Moreover, if the duration of data transmission between the local sensors and the fusion center is large, then the primary user may switch back to activity, thereby depriving the secondary network of a communication opportunity. In the cognitive radio setting, not only many terminals are needed for robust detection of primary activity, but also this detection should be done in the quickest way possible and under very strict constraints regarding the communication between the cognitive detectors and the fusion center.

Sequential detection is known to reduce the time of detection for a certain specified detection reliability (see, e.g., \cite{wald,poor,helstrom}). In contrast with block detection that operates on a fixed predetermined number of samples, the number of samples in sequential detection is a random variable. In addition, for the case of binary hypothesis testing, sequential detection requires two thresholds and not only one for operation. This allows for the decision to continue acquiring more samples to satisfy some detection performance measures. There has been considerable interest in applying sequential analysis to distributed or decentralized detection problems (see, e.g., \cite{decentralized_wald,ibrahim_varshney,thresholds,hashemi_decentralized_sequential,veeravalli,veeravalli2}). According to \cite{blum97}, distributed implementation of sequential detectors can have several forms. For instance, the sensors may forward their local decisions to the fusion center, which then runs a sequential test to obtain a global decision. On the other hand, the sensors themselves may perform sequential tests and cooperate together with or without a fusion center. In \cite{bandwidth_management}, the problem of distributed sequential detection in the presence of communication constraints is studied. An algorithm is developed for optimal rate or bandwidth distribution among detectors under a fixed bandwidth constraint. The work is based on the sequential probability ratio test (SPRT) of Wald \cite{wald}.

In the context of cognitive radio, distributed sequential detection has been investigated, for example, in \cite{kundargi_tewfik,alisayed,sequential_sensing_icc2010,tewfik_doubly_sequential_icc2010}. In the scheme proposed in \cite{alisayed}, each sensor computes the log-likelihood ratio (LLR) for its observations and reports it to the fusion center over a perfect reporting channel. The LLR's are accumulated sequentially at the fusion center till their sum is found sufficient to cross either of two predefined thresholds.  The thresholds are obtained on the basis of the desired global detection and false alarm probabilities. The authors also investigate the case of sequential detection with model uncertainties. In \cite{sequential_sensing_icc2010}, the detectors are divided into different sets according to the local signal-to-noise (SNR) of the sensing channel. The set with highest SNR sends to the fusion center first, followed by lower SNR until the fusion center can make a final decision. A doubly sequential test is proposed in \cite{tewfik_doubly_sequential_icc2010} in which both the local detectors and the fusion center operate sequential tests. The detectors have thresholds computed using Wald's approximations \cite{wald}. When the first sensor reaches a decision, it conveys its decision to the fusion center which is running a second sequential test for a global decision on channel availability. The authors make several suggestions pertaining to the termination criterion at the fusion center. For example, the fusion center stops when it has received a specified number of decisions in favor of a particular hypothesis. 

In addition to distributed sequential detection, we make use of the idea of ordered transmissions provided in \cite{sadler} for the purpose of energy efficient signal detection in wireless sensor networks (see also \cite{group_ordered}). By ordered transmissions, it is meant that the local sensor with the most reliable current measurement sends first its LLR value to the fusion center. In order for a sensor to know that it has the most reliable LLR value, there is no need for extensive information exchange. A timer backoff mechanism is suggested in \cite{sadler} to resolve this issue. Note that the sensor selection criterion is its instantaneous measurement quality and not the average SNR as in \cite{sequential_sensing_icc2010} for example. We discuss the scheme in detail in this paper. Suffice it now to mention that the proposed scheme in \cite{sadler} achieves the minimum average probability of error attained when all the sensor LLR's are used for a decision, and at the same time lowers the average number of transmissions needed to reach a decision. The scheme in \cite{sadler} is utilized in \cite{sadler_for_wsn} for sequential fusion in wireless sensor networks. It is certainly beneficial for a cognitive radio setting because it enhances the secondary throughput by quickly seizing the available spectral opportunities. 

In this paper, in order to reach a global decision at the secondary fusion center as quickly and reliably as possible, we adopt the ordered transmission scheme of \cite{sadler}. However, we impose a strict constraint on the maximum number of transmissions to the fusion center. We therefore extend the scheme to work for the case when the fusion center is to make a decision after receiving a maximum of $K$ observations from a total of $M$ local sensors. Assuming a slotted primary network where the primary user may switch activity every fixed amount of time, only few observations can be used within a time slot to make a decision regarding the availability of a transmission opportunity.  In other words, the decision regarding primary activity must be made soon enough to allow for channel access, and under the reasonable assumption of a narrowband cognitive control channel, only a few sensors can actually participate in the detection process.

Having imposed this constraint of a maximum number of reports from the local sensors, we devise a scheme where the decision thresholds employed at the secondary fusion center are computed using a dynamic programming framework. Given the LLR observations sent to the fusion center, the posterior probability of the channel being free is calculated and compared to the thresholds for a decision. We derive expressions for the likelihood functions of ordered LLR's and use these in both proposed schemes. For the dynamic programming scheme, a weighted sum of primary and secondary throughput objective function is constructed that also accounts for transmission costs, lost transmission opportunities and a penalty for collision with primary transmission. It is important to note that, in \cite{giannakis_sequential_cooperative}, dynamic programming is also employed to get the decision thresholds. However, it is assumed that all cognitive sensors would be able to report their energy measurements to the fusion center, where each detector senses one of the multi-band channels. 

In summary we make the following contributions in this paper. We use a cooperative sequential detection framework to address the tradeoff between throughput and detection reliability in cognitive radio systems taking into account the strict time limitation during which the decision must be made, which also imposes a limit on the number of sensors involved in the decision process. We employ an ordered LLR transmission scheme modifying the method of threshold determination in \cite{sadler} to account for the constraint on the number of reporting sensors. We also use dynamic programming to obtain the decision thresholds. The proposed method improves the throughput performance, which is the desired goal in the context of cognitive radios. We prove that if sensing has no cost, the optimal sequential decision is either to declare that the primary is idle or to continue sampling without ever deciding that the primary is active except possibly at the last decision stage. Finally, we address the impact of fading on the reporting channels linking the cognitive detectors to the fusion center.

The rest of the paper is organized as follows: Section \ref{sec:SysMdl} presents the system model of cooperative spectrum sensing in a cognitive radio network. Section \ref{sec:ord_scheme} discusses the previous work on ordered transmissions scheme, then discusses our proposed extension derivation. Section \ref{sec:seq} and \ref{sec:max_throughput} describe the proposed dynamic programming scheme. The effect of fading on the channels between the local detectors and the fusion center is explained in Section \ref{sec:fading_channels}. In Section \ref{sec:SimResults} we provide simulation results and elaborate on the difference in performance when both schemes are used. We conclude the paper in Section \ref{sec:conc}.

\section{System Model}
\label{sec:SysMdl}
\begin{figure}[htbp]
	\centering
		\includegraphics[width=0.50\textwidth]{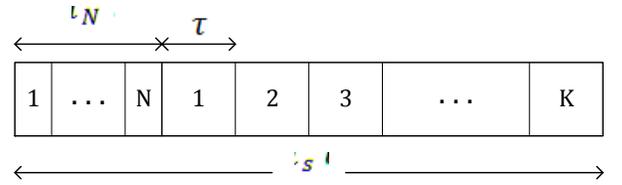}
	\caption{The primary time slot has a duration of $\tau_s$ units of time. Over $\tau_N$, each one of $M$ cognitive sensors takes $N$ samples at the beginning of the time slot. Afterwards, there are $K$ mini-slots, each of duration $\tau$, over which one LLR observation is transmitted from one cognitive sensor to the fusion center over a low-rate common control channel.}
	\label{fig:Time_slot_drawing}
\end{figure}

We consider a slotted primary system as shown in Figure \ref{fig:Time_slot_drawing}, where the primary activity, whether on or off, does not change during the time slot duration, $\tau_{\rm s}$. Primary activity switches independently from one slot to the next. There are $M$ cognitive sensors which take a number of measurements at the beginning of each time slot and compute a function of these measurements. A maximum of $K$ sensors among the $M$ sensors forward the results sequentially to a fusion center at which the final decision regarding primary activity is taken. We consider binary hypothesis testing at the fusion center with the following two hypotheses: 
\begin{center}

$H_0$: Sensed channel is free ; $H_1$: Sensed channel is busy    

\end{center}
The prior probabilities of each, denoted by $\pi_0$ and $(1-\pi_0)$, respectively, are assumed to be known. Observations from different sensors are conditionally independent given either hypothesis but can be non-identical. Let $X_i\left(n\right)$ be the received signal at the $i^{th}$ sensor at instant $n$, where $i = 1,2,...,M$. At each sensor $i$, $X_i\left(n\right)$ are independent given each hypothesis and are identically distributed. Assume that a total of $N$ samples are taken over a time duration $\tau_N$. Under the two hypotheses, $X_i\left(n\right)$ is given by

\begin{equation}
\begin{array}{l}

H_0 : X_i(n)=W_i(n)\:, \ \ \ \ \ \ \ \ \ \ \ \ \ \: n=1,2,...,N  \\
H_1 : X_i(n)=S_i(n)+W_i(n)\:, \ \ \ \; n=1,2,...,N

\end{array} 
\label{eq:xdef}
\end{equation}
where $W_i$ is additive white Gaussian noise (AWGN) having the same noise power, $\sigma^{2}$, at all sensors. Without loss of generality, the received primary signal $S_i\left(n\right)$ is assumed to be real zero-mean Gaussian random variable. The conditional probability distributions of $X_i\left(n\right)$ given $H_0$ and $H_1$ are described by $f_{X_i\left(n\right) }(x_n \mid H_0)$ and $f_{X_i\left(n\right)}(x_n \mid H_1)$, respectively, such that 
 
\begin{equation}
\begin{array}{l}

f_{X_i\left(n\right)}(x_n \mid H_0) \; \sim \; N(0,\sigma^{2})  \\
f_{X_i\left(n\right)}(x_n \mid H_1) \; \sim \; N(0,\sigma^{2}_{s_i}+\sigma^{2})

\end{array} 
\label{eq:xdef}
\end{equation}
where $\sigma^{2}_{s_i}$ is defined as the average received primary signal power at the $i^{th}$ local sensor, and is assumed to be fixed over a time slot and to change relatively slowly over time. The values of $\sigma^{2}_{s_i}$ are assumed to be known at the local sensors and are transmitted periodically after being quantized and coded to the fusion center on a low-rate common control channel that is not on the band being sensed. The quantization noise effect is assumed to be negligible. The LLR at the $i^{th}$ sensor is defined as 

\begin{equation}
Y_i = \sum_{n=1}^{N}\log \:\left[ \frac{f_{X_i\left(n\right) }(x_n\mid H_1)}{f_{X_i\left(n\right)}(x_n \mid H_0)}\right]
\end{equation}
This is the quantity that is computed by the local detector and reported to the fusion center sequentially as explained below.  Defining the local signal-to-noise ratio (SNR) as {\itshape $ \gamma_i = \frac{\textstyle{\sigma^{2}_{s_i}}}{\textstyle{\sigma^{2}}} $}, we can compute the LLR at the $i^{th}$ sensor as

\begin{eqnarray}
\label{LLR_at_sensor_i}
Y_i &=&  \frac{1}{2\sigma^{2}} \cdot \frac{\gamma_i}{\gamma_i+1} \displaystyle\sum\limits_{n=1}^N \mid X_i(n)\mid^2 \nonumber\\
&& - \; \log \; (1+\gamma_i)\:\frac{N}{2}
\end{eqnarray}
\noindent Note that under $H_0$ the summation $\sum_{n=1}^{N}\frac{\mid X_i\left(n\right)\mid^2}{\sigma^2}$ is a chi-square distribution with $N$ degrees of freedom. This is the same for the summation $\sum_{n=1}^{N}\frac{\mid X_i\left(n\right)\mid^2}{\sigma^2\left(1+\gamma_i\right)}$ under $H_1$. That is, the likelihood functions of $Y_i$ given $H_0$ or $H_1$ are shifted and scaled chi-square distributions with $N$ degrees of freedom.

The LLR values are quantized and coded using a fixed number of bits. As in the case of the local signal power, the quantization noise is assumed to be negligible. The information exchange between the cognitive sensors and the fusion center occurs on the common control channel. We initially assume that the transmission of the LLR values to the fusion center is perfect, then we consider the effect of fading in the channels between the local sensors and the fusion center in Section \ref{sec:fading_channels}. We assume that the procedure of seizing the control channel and sending the observation of one sensor requires an amount of time which we denote as $\tau$. Figure \ref{fig:Time_slot_drawing} illustrates the time durations $\tau_N$, $\tau$ and $\tau_s$. 

Since the transmission of each computed LLR value takes some time $\tau$, eliciting another measurement from the sensors, though improving the reliability of detection, wastes a duration of $\tau$ from the potential secondary transmission opportunity in case the primary is off. This causes a decrease in secondary throughput. In other words, we have a reliability-throughput tradeoff \cite{sensing_throughput_tradeoff} which we can control by allowing only the subset of the cognitive detectors with the most reliable observations to transmit their LLR's to the fusion center. This is implemented by having two thresholds at the fusion center. The LLR with maximum magnitude is transmitted first to the fusion center. A decision can be made, but if the metric is between the two thresholds, the second highest LLR in magnitude is transmitted to the fusion center and is combined with the first, and then a decision is attempted again. This continues until a decision in favor of $H_0$ or $H_1$ is made, or $K$ LLR's are accumulated at the fusion center. If $k$ sensors, $1\leq k \leq K$, are probed before a decision is reached, the time to make a decision is $\tau_N+k\tau$. Since the slot duration is $\tau_s$, this leaves $\tau_s-\tau_N-k\tau$ for transmission. Given the durations $\tau_N$, $\tau$ and $\tau_s$, it is obvious that $K$ must satisfy the inequality $\tau_s-\tau_N-K\tau \geq 0$.
  
Some final notes are in order regarding the operation of the secondary network. 
\begin{enumerate}[(a)]
\item The LLR values can be quantized to a few bits only with no considerable performance degradation when compared with the case where unquantized LLR's are used \cite{llr_quantizer}.
\item We assume a single hop network in which any broadcast by a secondary node is heard by the fusion center and all other nodes. This assumption is necessary such that when a node seizes the common channel to transmit its LLR, all other nodes become aware of its transmission. 
\item We assume that the sensors have nearly synchronized clocks. This issue and its practical aspects are discussed extensively in \cite{blum_ordering}.   
\end{enumerate}

\section{Ordered Transmissions Scheme}
\label{sec:ord_scheme}

In this section we describe the scheme in \cite{sadler}. A backoff timer is set at each of the $M$ sensors according to the magnitude of the locally computed LLR. Specifically, the timer is decreased as the absolute value of the sensor's LLR increases. Consequently, transmissions proceed such that the most confident observations and informative measurements are sent first to the fusion center. The fusion center then compares the accumulated sum of the received LLR's to two thresholds and makes one of three decisions accordingly; declare $H_0$, continue taking the next ranked observations, or declare $H_1$. Let $Y^{[m]}$ denote the LLR of rank $m=1,2,...,K $ received during the $m^{th}$ reporting mini-slot, where $m=1$ is the highest rank. For $1 \leq k \leq K$, the decision strategy is expressed as

\begin{equation}
{\rm If \,\,\,} \sum_{m=1}^{k}Y^{[m]} 
\begin{cases} 
 < t^{\left(L\right)}_k \ \ \ \ \ \ \ \ \ \ \ \ \ \ \ \text{Declare} \; H_0 \\
 \in (t^{\left(L\right)}_k,t^{\left(H\right)}_k) \ \ \ \ \ \ \ \text{Continue} \\
 > t^{\left(H\right)}_k  \ \ \ \ \ \ \ \ \ \ \ \ \ \ \ \text{Declare} \; H_1 
\end{cases}
\end{equation}
\noindent Note that the thresholds are generally time-dependent, i.e, their values depend on the stage index $k$. In order to force a stopping at $k=K$, $t^{\left({\rm L}\right)}_K=t^{\left({\rm H}\right)}_K$.

When the statistics evidently favor one hypothesis over the other, the decision is made and LLR transmissions from the local sensors stop. Otherwise, more observations are taken possibly till the end of the primary time slot. 
In \cite{sadler}, the data-dependent thresholds $t^{\left({\rm L}\right)}_k$ and $t^{\left({\rm H}\right)}_k$ are given by

\begin{eqnarray}
\label{sadler_thresholds}
t^{\left({\rm L}\right)}_k\;= \log\; (\frac{\pi_0}{1-\pi_0}) - \left(M-k\right) \; \big\vert Y^{[k]} \big\vert \nonumber \\
t^{\left({\rm H}\right)}_k= \log\; (\frac{\pi_0}{1-\pi_0}) + \left(M-k\right)  \; \big\vert Y^{[k]}\big\vert
\end{eqnarray}
\noindent where $M-k$ is the number of sensors that have not transmitted their LLR's yet at the ${k^{th}}$ stage, and $|Y^{[k]}|$ is the absolute value of the LLR received at stage $k$.  However, if this scheme is forced to take a decision in favor of either $H_0$ or $H_1$ at stage $K$, then it should work only for the case $K=M$, as can be seen by the fact that only $ t^{\left(L\right)}_M=t^{\left(H\right)}_M$. The transmission is ordered in terms of the absolute LLR value. That is, $|Y^{[m]}| < |Y^{[k]}|, \forall m > k$. Hence, $|Y^{[k]}|$ is greater than any LLR value transmitted later to the fusion center. It can be shown that this choice of thresholds ensure that this scheme has the same average probability of error as maximum a posteriori (MAP) procedure where all the LLR's are summed and compared to $\log\left(\pi_0/\left(1-\pi_0\right)\right)$ \cite{sadler}. The average probability of error when the guessed hypothesis, $\widehat{H}$, is not the true one is given by 

\begin{eqnarray}
\label{eq:avg_perror}
P_e &=& (1-\pi_0)\; Pr\left(\widehat{H}=H_0 \mid H_1\right)  \nonumber \\
   & & +\; \pi_0\; Pr\left(\widehat{H}=H_1 \mid H_0\right) 
\end{eqnarray} 
\noindent The advantage of this system is that the average number of transmissions needed to reach a decision is about half the total number of the sensors communicating with the fusion center \cite{sadler}. 

Here, and in our ICASSP paper \cite{CSS_ICASSP}, we consider extending the scheme in \cite{sadler} to work for $K \leq M$ case such that we only process the $K$ LLR's with highest magnitudes out of the $M$ LLR's. Given the sequence of observations $Y^{[1]}=y_1, Y^{[2]}=y_2, ..Y^{[K]}=y_K$, the optimal MAP block detector when the highest in magnitude $K$ out of $M$ LLR values are used has the following decision rule:
\begin{equation}
\label{block_detector}
\log \left[ \frac{f_{Y^{[1]}..Y^{[K]}\big\vert H_1}\left(y_1,y_2,..y_K \Big\vert H_1\right)}{f_{Y^{[1]}..Y^{[K]} \big\vert H_0}\left(y_1,y_2,..y_K \Big\vert H_0\right)} \right] \substack{ H_1\\ > \\ < \\ H_0} \log \frac{\pi_0}{1-\pi_0} 
\end{equation}
\noindent where $f_{Y^{[1]}..Y^{[K]}\big\vert H}\left(y_1,y_2,..y_K \big\vert H\right)$ is the joint density function of the magnitude-ordered LLR's given hypothesis $H$. 

In this section, we consider only the case of identical sensors. Let the probability density function of LLR value $y$ under hypothesis $H$ be given by $f_{Y}\left(y|H\right)$. Assuming the statistical independence of LLR's at various sensors, the joint density functions in (\ref{block_detector}) can be given by \cite{Order_Statistics}
\begin{equation}
\label{joint_dist}
\begin{split}
&f_{Y^{[1]}..Y^{[K]}\big\vert H}\left(y_1,y_2,..y_K \big\vert H\right)=\\
&f_Y\left(y_1 \big\vert H\right)..f_Y\left(y_K \big\vert H\right)\left({\rm Pr}\{\big\vert Y \big\vert \leq \big\vert y_K \big\vert \}\right)^{M-K}
\end{split}
\end{equation}
\noindent (See Appendix~\ref{sec:appendix2} for details).

In order to perform sequential detection, we compute the accumulated sum at the fusion center at stage $k$, which is given by $ \displaystyle \sum_{m=1}^{k}\log \:\left[ \frac{f_Y(y_m\mid H_1)}{f_Y(y_m\mid H_0)} \right]$. However, this accumulated sum can be replaced by $\sum_{m=1}^{k}y_m $, because as shown in Appendix~\ref{sec:appendix2} such processing will yield the same value of the LLR received at each stage $k$. This accumulated sum is then compared to two thresholds, $\hat{t}^{\left({\rm L}\right)}_k$ and $\hat{t}^{\left({\rm H}\right)}_k$, such that when it is greater than $\hat{t}^{\left({\rm H}\right)}_k$, $H_1$ is declared, and if it is smaller than $\hat{t}^{\left({\rm L}\right)}_k$, then $H_0$ is declared. Otherwise, if it does not cross either thresholds, the fusion center continues to accumulate the LLR's from the next ranked sensors. For all the stages with the exception of the $K^{th}$ stage, i.e., for $1 \leq k \leq K-1$, these thresholds are given by 
\begin{equation}
\begin{split}
\hat{t}^{\left({\rm L}\right)}_k= &\log\: \frac{\textstyle{\pi_0}}{\textstyle{1-\pi_0}} \\ 
&- \left(K-k\right) \: \big\vert y_k \big\vert -(M-K)\displaystyle\max_{0\leq y \leq |y_k|} \rho\left(y\right) \\
\hat{t}^{\left({\rm H}\right)}_k= &\log\: \frac{\textstyle{\pi_0}}{\textstyle{1-\pi_0}} \\
&+ \left(K-k\right) \: \big\vert y_k \big\vert -(M-K)\displaystyle\min_{0\leq y \leq |y_k|} \rho\left(y\right)
\label{mod_sadler_thresholds}
\end{split}
\end{equation}
\noindent where for LLR $Y$
\begin{equation}
\begin{split}
\rho\left(y\right)&=\log \frac{{\rm Pr}\{\big\vert Y \big\vert \leq y \big\vert H_1\}}{{\rm Pr}\{\big\vert Y \big\vert \leq y \big\vert H_0\}}\\
&=\log \frac{\int_{-y}^{y}f_{Y}\left(w|H_1\right)dw}{\int_{-y}^{y}f_{Y}\left(w|H_0\right)dw}
\end{split}
\end{equation}
is a correction term to (\ref{sadler_thresholds}) that appears at each stage $k$ to account for the probability that all the other $(M-K)$ sensors would have LLR values less than $|y_k|$. Since it is not a monotonic function, we need to span over the range from $0<y<|y_k|$ and choose the maximum and the minimum values of $\rho(y)$. At the end of the slot, i.e., $k=K$, 
\begin{equation}
\hat{t}^{\left({\rm H}\right)}_K=\hat{t}^{\left({\rm L}\right)}_K=\log\: \frac{\textstyle{\pi_0}}{\textstyle{1-\pi_0}}-(M-K)\rho(y_K)  
\end{equation}

We now prove that using the thresholds in (\ref{mod_sadler_thresholds}) ensures that the average probability of error achieved by the sequential scheme using the above thresholds is the same as the minimum average probability of error attained by a MAP detector operating on $K$ out of $M$ measurements with the highest absolute magnitude. Consider the situation in which the accumulated LLR exceeds $\hat{t}^{\left({\rm H}\right)}_k$.
\begin{equation}
\sum_{m=1}^{k} y_{m} \geq \hat{t}^{\left({\rm H}\right)}_k
\end{equation} 
\noindent Note that 
\begin{equation}
\begin{split}
\left(K-k\right) \Big\vert y_k\Big\vert \geq \sum_{m=k+1}^{K} \Big\vert y_m\Big\vert \geq -\sum_{m=k+1}^{K} y_m
\nonumber
\end{split}
\end{equation}
\noindent Since $|y_k|\geq |y_K|$, minimizing over the domain $0\leq y \leq |y_k|$ achieves at least the same minimum when the domain $0\leq y \leq |y_K|$ is used. Therefore,
\begin{equation}
\begin{split}
\min_{0\leq y \leq |y_k|}\rho\left(y\right)&=\min_{0\leq y \leq |y_k|}\log \frac{{\rm Pr}\{\big\vert Y \big\vert \leq y \Big\vert H_1\}}{{\rm Pr}\{\big\vert Y \big\vert \leq y \Big\vert H_0\}}\\
& \leq \min_{0\leq y \leq |y_K|}\log \frac{{\rm Pr}\{\big\vert Y \big\vert \leq y \Big\vert H_1\}}{{\rm Pr}\{\big\vert Y \big\vert \leq y \Big\vert H_0\}}\\
& \leq \log \frac{{\rm Pr}\{\big\vert Y \big\vert \leq \big\vert y_K \big\vert \Big\vert H_1\}}{{\rm Pr}\{\big\vert Y \big\vert \leq \big\vert y_K \big\vert \Big\vert H_0\}}
\nonumber
\end{split}
\end{equation}
\begin{equation}
\begin{split}
-\left(M-K\right)&\min_{0\leq y \leq |y_k|}\rho\left(y\right) \geq \\
&-\left(M-K\right)\log \frac{{\rm Pr}\{\big\vert Y \big\vert \leq \big\vert y_K \big\vert \Big\vert H_1\}}{{\rm Pr}\{\big\vert Y \big\vert \leq \big\vert y_K \big\vert \Big\vert H_0\}}
\nonumber 
\end{split}
\end{equation}
\noindent Hence, when $\sum_{m=1}^{k} y_m$ exceeds $\hat{t}^{\left({\rm H}\right)}_k$, it also exceeds 
\begin{equation}
\log \frac{\pi_0}{1-\pi_0}-\sum_{m=k+1}^{K} y_m -\left(M-K\right)\log \frac{{\rm Pr}\{\big\vert Y \big\vert \leq \big\vert y_K \big\vert \Big\vert H_1\}}{{\rm Pr}\{\big\vert Y \big\vert \leq \big\vert y_K \big\vert \Big\vert H_0\}}
\nonumber
\end{equation}
\noindent This implies that 
\begin{equation}
\sum_{m=1}^{K} y_m +\left(M-K\right)\log \frac{{\rm Pr}\{\big\vert Y \big\vert \leq \big\vert y_K \big\vert \Big\vert H_1\}}{{\rm Pr}\{\big\vert Y \big\vert \leq \big\vert y_K \big\vert \Big\vert H_0\}} \geq \log \frac{\pi_0}{1-\pi_0}
\nonumber
\end{equation}
\noindent which is the MAP rule for deciding $H_1$ based on a block of best $K$ among $M$ sensors. The case of deciding $H_0$ can be proved similarly. The benefit reaped from using the sequential scheme is that the average number of observations needed to make a decision is close to $\frac{K}{2}$, which has been showed via simulations. In the next section, we adopt a method for determining the thresholds that allows a tradeoff between the average probability of error and the speed of reaching a decision concerning primary activity.

\section{Proposed Ranked Sequential Scheme}
\label{sec:seq}

In this section, we use dynamic programming to calculate the thresholds for sequential detection. Specifically, we employ the backward induction technique which provides the optimal action to be taken in order to minimize the overall decision cost \cite{Bertsekas_DP}. In the analysis below we assume that the fusion center knows the statistics of all detectors, which are updated periodically over the common control channel and change relatively slowly over time as mentioned in Section \ref{sec:SysMdl}.  

Since we adopt the ordered transmission scheme and assume perfect reporting, if an LLR value is received, all subsequent LLR's would have values with a lesser magnitude. In other words, if the value of the LLR received at stage $k-1$ is equal to $\alpha$, then the probability ${\rm Pr}\left(|Y^{[m]}|>|\alpha|\right)=0$, $\forall m \geq k$. Recall that $Y_{m}$ is the LLR from the $m^{th}$ sensor, $1 \leq m \leq M$, whereas $Y^{[m]}$ is the LLR with the $m^{th}$ highest magnitude that is transmitted to the fusion center at the $m^{th}$ stage with $m=1,2,..K$. 

Let $\pi_k$ be the probability of the channel being idle at stage $k$ given the sequence of observations $Y^{[1]}=y_1, Y^{[2]}=y_2, ..Y^{[k]}=y_k$. Probability $\pi_k$ can be obtained recursively as follows:
\begin{equation}
\label{recursive}
\begin{split}
&\pi_k={\rm Pr}\left(H_0|y_1,y_2,..y_k\right) =\\
& \frac{f_{Y^{[k]}|Y^{[1]}..Y^{[k-1]},H_0}\left(y_k|y_1..y_{k-1},H_0\right){\rm Pr}\left(H_0|y_1..y_{k-1}\right)}{\sum\limits_{r=0,1}f_{Y_{[k]}|Y^{[1]}..Y^{[k-1]},H_r}\left(y_k|y_1..y_{k-1},H_r\right){\rm Pr}\left(H_r|y_1..y_{k-1}\right)} \\
& = \frac{f_{Y^{[k]}|Y^{[k-1]},H_0}\left(y_k|y_{k-1},H_0\right)\pi_{k-1}}{\sum\limits_{r=0,1}f_{Y^{[k]}|Y^{[k-1]},H_r}\left(y_k|y_{k-1},H_r\right)\left(r+\left(1-2r\right)\pi_{k-1}\right)}
\end{split}
\end{equation} 
\noindent where in the last step, we use the Markovian property induced by ordered transmissions that $Y^{[k]}$ is independent of $Y^{[1]}, Y^{[2]}, ..Y^{[k-2]}$ given $Y^{[k-1]}$ and either hypothesis (Theorem 2.4.3, \cite{Order_Statistics}).

 What is needed for backward induction is the conditional probability of $Y^{[m]}$ given $Y^{[m-1]}$ under both $H_0$ and $H_1$. Assume that the probability density function of LLR value $Y_{k}$ under hypothesis $H$ (either $H_0$ or $H_1$) is given by $f_{Y_{k}}\left(y|H\right)$, which is a scaled and shifted chi-square distribution with $N$ degrees of freedom as mentioned in Section \ref{sec:SysMdl}. Define $\beta_{k,b,H}={\rm Pr}\left\{\left[|Y_k|>|b|\right]\mid H\right\}$. The value of $\beta_{k,b,H}$ can be readily computed knowing the likelihood functions of $Y_k$. For non-identical sensors, the joint distribution of $Y^{[m]}$ and $Y^{[m-1]}$ conditioned on hypothesis $H$ when $-|\gamma|<\alpha<|\gamma|$ and $m \geq 2$ is given by:
\begin{equation}
\begin{split}
f_{Y^{\left[m\right]},Y^{\left[m-1\right]}}&\left(\alpha,\gamma|H\right)= \sum_{k=1}^{M} \sum_{j \neq k}^{M} f_{Y_k}\left(\alpha|H\right)f_{Y_{j}}\left(\gamma|H\right) \cdot \\
& \sum_{S^{m-2}_{k,j}} \prod_{v \neq k,j} \left(\beta_{v,\gamma,H}\right)^{x_{v}}\left(1-\beta_{v,\alpha,H}\right)^{\left(1-x_{v}\right)}
\end{split}
\label{joint}
\end{equation}
\noindent where $S^{m-2}_{k,j}$ is a set of $m-2$ sensors chosen from the $M$ sensors with sensors $k$ and $j$ excluded. The number of sets $S^{m-2}_{k,j}$ is $M-2 \choose m-2$. Parameter $x_{v}$ is equal to unity when $v \in S^{m-2}_{k,j}$ and zero otherwise. $f_{Y^{\left[m\right]},Y^{\left[m-1\right]}}\left(\alpha,\gamma \mid H\right)=0$ if $|\alpha|>|\gamma|$. Note that (\ref{joint}) is derived as follows. The joint distribution between $Y^{[m]}$ and $Y^{[m-1]}$ involves a summation over all possible sensor pairs. Since, given a particular hypothesis, the LLR's from the local sensors are independent, we get the product $f_{Y_k}\left(\alpha \mid H \right)f_{Y_{j}}\left(\gamma \mid H\right)$. For $m \geq 2$ and due to the ordered nature of LLR transmission, we have $m-2$ sensors with absolute LLR values exceeding $Y^{[m-1]}=\gamma$. The rest of the sensors should have absolute LLR values that are less than $Y^{[m]}=\alpha$.  In the case of identical sensors, $f_{Y_k}\left(y \mid H\right)=f_{Y}\left(y \mid H\right)$ and $\beta_{k,b,H}=\beta_{b,H}$,

\begin{equation}
\begin{split}
f&_{Y^{\left[m\right]},Y^{\left[m-1\right]}}\left(\alpha,\gamma \mid H\right)= M\left(M-1\right) \cdot \\
& f_{Y}\left(\alpha \mid H\right)f_{Y}\left(\gamma \mid H \right) {{M-2} \choose {m-2}}  \left(\beta_{\gamma,H}\right)^{m-2}\left(1-\beta_{\alpha,H}\right)^{M-m}
\end{split}
\end{equation}

In order to obtain the conditional distribution $f_{Y^{\left[m\right]}|Y^{\left[m-1\right]}}\left(\alpha|\gamma,H\right)$, we need the distribution $f_{Y^{[m-1]}}\left(\gamma|H\right)$ because
\begin{equation}
\label{cond_dist_of_llrs}
f_{Y^{\left[m\right]}|Y^{\left[m-1\right]}}\left(\alpha|\gamma,H\right)=\frac{f_{Y^{\left[m\right]},Y^{\left[m-1\right]}}\left(\alpha,\gamma \mid H\right)}{f_{Y^{[m-1]}}\left(\gamma|H\right)}
\end{equation}
\noindent Distribution $f_{Y^{[m-1]}}\left(\gamma|H\right)$ can be readily obtained as
\begin{equation}
\label{pdf_of_ranked_llr}
\begin{split}
& f_{Y^{[m-1]}}\left(\gamma|H\right)= \\ 
& \sum_{k=1}^{M} f_{Y_{k}}\left(\gamma|H\right)
\sum_{S^{m-2}_{k}} \prod_{v \neq k} \left(\beta_{v,\gamma,H}\right)^{x_{v}}\left(1-\beta_{v,\gamma,H}\right)^{\left(1-x_{v}\right)}
\end{split}
\end{equation}
\noindent where $S^{m-2}_{k}$ is a set of $m-2$ sensors chosen from the $M$ sensors with sensor $k$ excluded. The number of sets $S^{m-2}_{k}$ is $M-1 \choose m-2$. Parameter $x_{v}$ is equal to unity when $v \in S^{m-2}_{k}$ and zero otherwise. In the case of identical sensors, $\beta_{k,\gamma,H}=\beta_{\gamma,H}$, 

\begin{equation}
\begin{split}
& f_{Y^{[m-1]}}\left(\gamma|H\right)= \\ 
& M f_{Y}\left(\gamma|H\right) {M-1 \choose {m-2}} \left(\beta_{\gamma,H}\right)^{m-2}\left(1-\beta_{\gamma,H}\right)^{M-m+1}
\end{split}
\end{equation}

We construct now the cost-to-go function which is used to obtain the optimal policy in optimal stopping problems \cite{optimal_stopping_book}. The optimal policy here is comprised of the optimal thresholds to be used at each stage to determine whether to stop and make a decision regarding the channel occupancy, or continue obtaining more LLR measurements. Assume that the cost of deciding \textit{i} at stage $k$ when \textit{j} is the true hypothesis is $\lambda^k_{ij}$, and the cost to continue taking observations is $c \geq 0$. Define $J^{\left(K\right)}_{k}$ as the minimum cost-to-go at stage $k$ of the finite horizon $K$.
\begin{equation}
\begin{split}
J^{\left(K\right)}_{k}(\pi_k,Y^{[k]}) &= \text{min} \bigg\{ \lambda^k_{00}\;\pi_k + \lambda^k_{01}\;(1-\pi_k),\\
& \lambda^k_{10}\;\pi_k + \lambda^k_{11}\;(1-\pi_k),\\
& c + \mathbb{E}_{Y^{[k+1]} \vert Y^{[k]}}\left[J^{\left(K\right)}_{k+1}(\pi_{k+1},Y^{[k+1]})\big\vert Y^{[k]}\right] \bigg\}
\label{minimum_cost}
\end{split}
\end{equation}
\noindent where $\pi_{k}$ and $\pi_{k+1}$ are related through the expression in (\ref{recursive}) with $k$ replaced by $k+1$. The first term in (\ref{minimum_cost}) is the cost of deciding $H_{0}$, whereas the second term is the cost of deciding $H_1$. These correspond to the channel decided to be free or busy, respectively. The third term in (\ref{minimum_cost}) is interpreted as the expected cost when the fusion center decides that it should continue taking more observations from local sensors. The conditional expectation is given by
\begin{equation}
\begin{split}
&\mathbb{E}_{Y^{[k+1]}\vert Y^{[k]}}\left[J^{\left(K\right)}_{k+1}(\pi_{k+1},Y^{[k+1]})\big\vert Y^{[k]}\right] = \\
& \int_{-\vert Y^{[k]}\vert}^{\vert Y^{[k]}\vert} {J^{\left(K\right)}_{k+1}(\pi_{k+1},y)f_{Y^{[k+1]} \big\vert Y^{[k]}}\left(y \big\vert Y^{[k]}\right)\;dy}
\end{split}
\end{equation}
and
\begin{equation}
\begin{split}
&f_{Y^{[k+1]}|Y^{[k]}}\left(y \big\vert Y^{[k]}\right)=f_{Y^{[k+1]}|Y^{[k]},H_0}\left(y \big\vert Y^{[k]},H_0\right)\pi_{k} \\
&+f_{Y^{[k+1]}|Y^{[k]},H_1}\left(y \big\vert Y^{[k]},H_1\right)\left(1-\pi_{k}\right)
\end{split}
\end{equation}

The parameter $c$ represents the tradeoff between the time taken till a decision is made and the average probability of error. As $c$ increases, the fusion center becomes more likely to favor one of the two hypotheses in a shorter time using only a few LLR's from the cognitive detectors. At the last stage, i.e., when $k=K$, the fusion center has two choices only; to declare $H_0$ or $H_1$. This allows backward induction from the last stage
\begin{equation}
\begin{split}
J^{\left(K\right)}_{K}(\pi_K,Y^{[K]}) = \text{min}\; \bigg\{ & \lambda^K_{00}\;\pi_K + \lambda^K_{01}\;\left(1-\pi_K \right), \\
& \lambda^K_{10}\;\pi_K + \lambda^K_{11}\;\left(1-\pi_K \right)  \bigg\}
\label{cost_T}
\end{split}
\end{equation}
\noindent Note that there is no actual dependence of $J^{(K)}_K$ on $Y^{[K]}$ because there is no more sampling after the $K^{th}$ stage. $J^{\left(K\right)}_{K-1}(\pi_{K-1},Y^{[K-1]})$ can be obtained using (\ref{minimum_cost}) for all values of $Y^{[K-1]}$. The process can then be repeated to obtain the thresholds and optimal decisions. It is clear that this is a computationally extensive task due to the dependence on last observation. In particular, the thresholds at each stage would be a function of the value of the LLR sent in the previous stage. To reduce the complexity, we propose here to use the distributions of $Y^{[k]}$ without conditioning. This makes the problem considerably more manageable. We have verified through simulations that at least for a few number of sensors dropping this dependency has a negligible effect on the objective function. Dropping the explicit dependence on $Y^{[k]}$, the dynamic program now acquires the form
\begin{equation}
\begin{split}
J^{\left(K\right)}_{k}(\pi_k) = \text{min}\; \bigg\{ & \lambda^k_{00}\;\pi_k + \lambda^k_{01}\;\left(1-\pi_k \right),\\
& \lambda^k_{10}\;\pi_k + \lambda^k_{11}\;\left(1-\pi_k \right),\\
& c + \mathbb{E}_{Y^{[k+1]}}\left[J^{\left(K\right)}_{k+1}(\pi_{k+1})\right] \bigg\}
\label{min_cost_approx}
\end{split}
\end{equation}
\noindent such that
\begin{equation}
\begin{split}
\mathbb{E}_{Y^{[k+1]}}\left[J^{\left(K\right)}_{k+1}(\pi_{k+1})\right] = \int {J^{\left(K\right)}_{k+1}(\pi_{k+1})f_{Y^{[k+1]}}\left(y\right)\;dy}
\end{split}
\end{equation}
\noindent with
\begin{equation}
f_{Y^{[k+1]}}\left(y\right)=\pi_{k}f_{Y^{[k+1]}}(y\mid H_0)+{(1-\pi_{k})}f_{Y^{[k+1]}}(y \mid H_1)
\end{equation}
\begin{equation}
\label{rec_pi_approx}
\pi_{k+1} = \frac{\pi_{k}f_{Y^{[k+1]}}\left(y \mid H_0\right)}{\pi_{k}f_{Y^{[k+1]}}(y\mid H_0)+{(1-\pi_{k})}f_{Y^{[k+1]}}(y \mid H_1)} 
\end{equation}
\noindent Finally, $f_{Y^{[k+1]}}(y\mid H)$ can be computed using an expression similar to (\ref{pdf_of_ranked_llr})
\begin{equation}
\label{pdf_of_ranked_llr}
\begin{split}
& f_{Y^{[k+1]}}\left(y|H\right)= \\ 
& \sum_{r=1}^{M} f_{Y_{r}}\left(y|H\right)
\sum_{S^{k}_{r}} \prod_{v \neq r} \left(\beta_{v,y,H}\right)^{x_{v}}\left(1-\beta_{v,y,H}\right)^{\left(1-x_{v}\right)}
\end{split}
\end{equation}

Note that the optimal thresholds and decisions can be calculated offline so long as the relevant system parameters are fixed. When the fusion center gets the LLR observations, it uses the pre-computed thresholds to decide whether to decide in favor of $H_0$ or $H_1$, or to request the transmission of one more LLR value from sensors. The calculation should be redone when any of the parameters, such as the local SNR's at the cognitive detectors, change significantly.  
\section{Performance Optimization}
\label{sec:max_throughput}

We now provide expressions for costs $\lambda_{ij}^{k}$ introduced in Section \ref{sec:seq}. Recall that $\lambda_{ij}^{k}$ is the cost to decide $i$ when $j$ is true at the $k^{th}$ mini-slot of duration $\tau$. Our objective is to optimize the system performance with a focus on the achievable primary and secondary throughput. Specifically, when we deal with throughput, we mean a weighted sum of primary and secondary throughputs where a factor $\omega$, $0\leq\omega\leq1$, is multiplied by primary throughput and $1-\omega$ is used to weight the secondary throughput. The closer $\omega$ to unity, the more emphasis we put on the primary rate. This implies more protection for the primary link against interference and service interruption. Parameter $\omega$ should be chosen by the primary network to satisfy the transmission and rate requirements of primary users. 

Define $\eta_p$ and $\eta_s$ to be the probability of correct reception of the primary and secondary signals in the presence of receiver noise only, respectively. Similarly, define $\delta_p$ and $\delta_s$ to be the probability of correct reception of the primary and secondary signals in the presence of noise in addition to interference from the other user due to concurrent transmission. Given that $R_p$ and $R_s$ are the primary and secondary rates of transmission, respectively, the costs of different decisions are then given by
\begin{equation}
\label{eq:lambda00}
\begin{split}
\lambda_{00}^{k} = &- (1-\omega)\:R_s \eta_s \frac{\left(\tau_s-\tau_N-k\tau\right)}{\tau_s}+ \\
& e_{st}\;\frac{\left(\tau_s-\tau_N-k\tau\right)}{\tau_s}
\end{split}
\end{equation}
\noindent where $e_{st}$ is the transmission cost or expended energy for the secondary terminal if it transmits for the whole slot duration. This cost is given in rate units to be compatible with the first term in (\ref{eq:lambda00}).
\begin{equation}
\label{eq_lambda01}
\begin{split}
\lambda_{01}^{k} =&  -\omega R_p\:\delta_p -\left(1-\omega\right)R_s \delta_s \frac{\left(\tau_s-\tau_N-k\tau\right)}{\tau_s} +\\
& e_{pt} + e_{st}\;\frac{\left(\tau_s-\tau_N-k\tau\right)}{\tau_s}+ P
\end{split}
\end{equation}
\noindent where $e_{pt}$ is the transmission cost or expended energy for the primary terminal expressed in units of rate, and $P$ is a penalty term for collision with primary user.
\begin{equation}
\label{eq:lambda10}
\lambda_{10}^{k}=L_{\rm f}
\end{equation}
\noindent where $L_{\rm f}$ is the cost of losing the opportunity to access the spectrum when the channel is idle. This term can be positive to account for other negative consequences for remaining idle than the throughput being zero. The failure to transmit may increase the transmission delay for a delay-sensitive application, or cause packet loss due to queue overflow, etc. 
\begin{equation}
\label{eq:lambda11}
\lambda_{11}^{k}= - \omega R_p \eta_p + e_{pt}+L_{\rm b} 
\end{equation}
\noindent where $L_{\rm b}$ is the cost of losing the opportunity to access the spectrum when the channel is busy. All the costs $e_{st}$, $e_{pt}$, $L_{\rm f}$, $L_{\rm b}$ and $P$ are nonnegative. 

The first term of expression (\ref{eq:lambda00}) shows the gain of the secondary user when an idle channel is correctly detected. It depends on the weight assigned to the secondary user, the time remaining from the time slot $\tau_s$ after making the decision, and the probability of correct detection in the absence of primary interference. Similarly, the first term of (\ref{eq:lambda11}) is the primary throughput in the case of correct detection by the secondary users. The first two terms in (\ref{eq_lambda01}) represent the possibility of correct reception in spite of the occurrence of a collision due to mis-detection. The interference survival probabilities $\delta_p$ and $\delta_s$ would be typically small in value.

Recall that parameter $c \geq 0$ is the cost of taking one more observation. Given the aforementioned performance metrics, $c$ can be understood as the cost, in units of rate, of the transmission phase in which the secondary fusion center solicits another LLR measurement and receives it from the appropriate local sensor. 

\subsection{Throughput Maximization}

If we adopt the weighted sum throughput maximization objective, then setting all the costs, including $c$, to zero except for the throughput terms, declaring $H_1$ at any stage in the slot except for the last stage might cause losing the opportunity to access the spectrum by the cognitive radio device. Instead, the fusion center should solicit more observations, for a possible declaration of a free channel before the end of the slot, which consequently increases the normalized secondary throughput. Hence, rather than two thresholds as in classical sequential detection, we can just have one threshold and two decisions: to declare $H_0$ and assign one of the cognitive terminals to transmit over the probed channel, or to ask for one more observation if less than $K$ LLR observations have been collected. Consequently, the objective cost-to-go at stage $k$ can be expressed as 
\begin{equation}
\begin{split}
J^{\left(K\right)}_{k}(\pi_k) = \text{min}\; \bigg\{ & \lambda^k_{00}\;\pi_k + \lambda^k_{01}\;\left(1-\pi_k \right),\\
& \mathbb{E}_{Y_{k+1}}\left[J^{\left(K\right)}_{k+1}(\pi_{k+1})\right] \bigg\}
\label{min_cost}
\end{split}
\end{equation}
\noindent where $\lambda^k_{00}$ and $\lambda^k_{01}$ are given by (\ref{eq:lambda00}) and (\ref{eq_lambda01}) with $e_{st}=e_{pt}=0$. The above expression is valid for $k<K$. At the last stage where $k=K$, we have
\begin{equation}
\begin{split}
J^{\left(K\right)}_{K}(\pi_K) = \text{min}\; \bigg\{ & \lambda^K_{00}\;\pi_K + \lambda^K_{01}\;\left(1-\pi_K \right),\\
& \lambda^K_{10}\;\pi_K + \lambda^K_{11}\;(1-\pi_K) \bigg\}
\label{min_costx}
\end{split}
\end{equation}
\noindent where $\lambda^K_{10}$ and $\lambda^K_{11}$ are given by (\ref{eq:lambda10}) and (\ref{eq:lambda11}) with $k=K$, $e_{st}=e_{pt}=0$ and $L_{\rm b}=L_{\rm f}=0$. We actually can show that starting with the two-threshold case, if we set $c=0$, $e_{st}=e_{pt}=0$ and $L_{\rm b}=L_{\rm f}=0$, the two-threshold case would converge to the one-threshold case as the lower threshold would always be set to $0$ except at the very last stage. The proof is given in Appendix~\ref{sec:appendix3}. Note that if there are multiple channels the two-threshold case would not necessarily converge to the the one-threshold case because declaring $H_1$ would be equivalent to switching to another channel in order to search for a transmission opportunity.  


\section{Fading Channels Between Sensors and Fusion Center}
\label{sec:fading_channels}

In this section, and instead of the previously assumed perfect reporting channels, we consider fading channels between the local detectors and the fusion center. The statistical distribution of the channel gain between the $i^{th}$ sensor and the fusion center is given by $f_{g_i}\left(x\right)$, possibly different for different sensors. We assume that the instantaneous values of the channel gains are known by the sensors and the fusion center. These values are fixed over a duration $T_{\rm c}$ which is multiples of the primary slot duration $\tau_s$. Note that if the channel between a sensor and a fusion center is in deep fade, then even if the sensor has a reliable observation, it may not be able to transfer it correctly to the fusion center within the mini-slot duration $\tau$. For the transmission from the $i^{th}$ sensor to the fusion center to be decodable, the rate of transmission should be lower than the link capacity which, for a fixed and known channel gain $g_i$ at the transmitter and additive white Gaussian noise at the receiver, is given by
\begin{equation}
\label{capacity_sensor_fc}
C_i=W\log\left(1+\frac{P_i g_i}{\sigma_{\rm f}^2}\right)
\end{equation}
\noindent where $W$ is the reporting channel bandwidth, $\sigma_{\rm f}^2$ is the noise variance at the receiver of the fusion center, and $P_i$ is the transmitted power by the $i^{th}$ sensor. We consider the case where a local sensor transmits with a rate $r_i$ given by
\begin{equation}
\label{rate_sensor_fc}
r_i=W\log\left(1+\frac{P_i g_i}{\Gamma_i \sigma_{\rm f}^2}\right)
\end{equation}
\noindent where $\Gamma_i>1$ is the signal-to-noise ratio gap to capacity \cite{Digital_Communication_Lee_Ch8}. This factor accounts for the practical limitations that forces transmission at a rate below link capacity for correct reception. Assume that a number $b$ of information bits needs to be transferred to the fusion center. This would include the quantized LLR values with high resolution to justify the neglect of quantization noise. The amount of time needed for this transfer conditioned that the fusion center would be able to decode the transmission correctly is given by $b/r_i$. We consider that this transmission should take no longer than $\tau_{\rm b}$ which is obviously smaller than $\tau$. In other words, the $i^{th}$ sensor would be able to send its LLR value to the fusion center in time $\tau_b$ or less if
\begin{equation}
\frac{b}{W\log\left(1+\frac{P_i g_i}{\Gamma_i \sigma_{\rm f}^2}\right)} \leq \tau_{\rm b}
\end{equation}
\noindent This can be written as
\begin{equation}
\label{transmission_condition}
g_i \geq \frac{\Gamma_i \sigma_{\rm f}^2}{P_i}\left(2^{\frac{b}{W\tau_b}}-1\right)
\end{equation} 
\noindent Let $\bar{g}_i$ be the right-hand-side of inequality (\ref{transmission_condition}). The probability then that a sensor would report its LLR measurement to the fusion center is given by 
\begin{equation}
\delta_i=\int_{\bar{g}_i}^{\infty} f_{g_i}\left(x\right)\, dx
\end{equation}

Over a period of $T_{\rm c}$ only a fraction of the local detectors would participate in the reporting process to the fusion center. Since the channels are assumed to be known at the sensors, each sensor can decide whether it should participate in the next sensing epoch. The fusion center also, knowing the channels, can know the subset of sensors that would be involved in the next sensing events that span a time duration of $T_{\rm c}$. Our analysis in the previous sections applies but with a number of sensors equal to those whose channels satisfy (\ref{transmission_condition}). The average number of sensors participating in the sensing is simply $\sum_{i=1}^{M}\delta_i$. The probability that the number of sensors participating in sensing is equal to $\bar{M}$, where $0\leq \bar{M}\leq M$, is given by
\begin{equation}
\sum_{S_{\bar{M}}}\prod_{i=1}^{M} \delta_i^{x_i}\left(1-\delta_i\right)^{1-x_i}
\end{equation}
\noindent where $S_{\bar{M}}$ is a set of $\bar{M}$ sensors from $M$ sensors. The number of elements in set $S_{\bar{M}}$ is $M \choose \bar{M}$. Factor $x_i=1$ if sensor $i$ belongs to the set $S_{\bar{M}}$ and zero otherwise. For symmetric channels between the sensors and the fusion center, $\delta_i=\delta$ for all $i$ and the probability of $\bar{M}$ sensors participating is
\begin{equation}
{M \choose \bar{M}}\delta^{\bar{M}}\left(1-\delta\right)^{M-\bar{M}}
\end{equation}
\noindent The average number of sensors participating in the sensing and reporting process is, hence, equal to $M\delta$. This means that the performance of the network with $M$ sensors would, on average, be equal to the performance of a network with non-fading reporting channels, albeit with $M\delta$ sensors. 
\section{Simulation Results}
\label{sec:SimResults}

In this section, we investigate via numerical simulations the performance of our proposed schemes. Unless otherwise stated, the simulation parameters are as follows. The noise variance at each detector $\sigma^{2} = 1$ and the transmission of each observation takes $\tau=0.1$ units of time. The time taken at the beginning of a time slot with duration $\tau_s=1$ to collect $N$ observations is $\tau_N=2\tau$. This leaves $K=8$ stages in the slot in which the fusion center takes the ordered LLR observations. We consider the case of equal prior probabilities of both hypotheses, i.e., $\pi_0 =0.5$, and equal local signal to noise ratio over the channels between the local sensors and the primary user, $\sigma^{2}_{s_i}=2,\; i=1,2,...,M$. Each sensor takes $N=3$ samples in the duration $\tau_N$. The cost, $c$, to continue without deciding on one of the two hypotheses is set to $0.0001$. 

In order to minimize the average probability of error in equation (\ref{min_cost_approx}), we use the decision costs:  $\lambda^k_{ij}=1$ when $i \neq j $ and $\lambda^k_{ij}=0$ when $i = j$. Cost $\lambda^k_{ii}=0$ means that no cost is incurred if the decision at stage $k$ is the true hypothesis. Moreover, to study the objective of maximizing the weighted sum throughput, we use the expressions (\ref{eq:lambda00})-(\ref{eq:lambda11}) for the costs. In the simulations, the transmission costs for both the primary and secondary users, $e_{pt}$ and $e_{st}$, are set to $0$, as well as the penalty factor $P$. Furthermore, the costs of losing the opportunity to access the spectrum when it is idle or busy, $L_f$ and $L_b$, are assumed to be $0$. The transmission success probabilities are $\eta_p=1$, $\eta_s=1$, $\delta_p=0$ and $\delta_s=0$. 

\subsection{Two-Threshold Based Method }

\begin{figure}[htbp]
	\centering
		\includegraphics[width=0.50\textwidth]{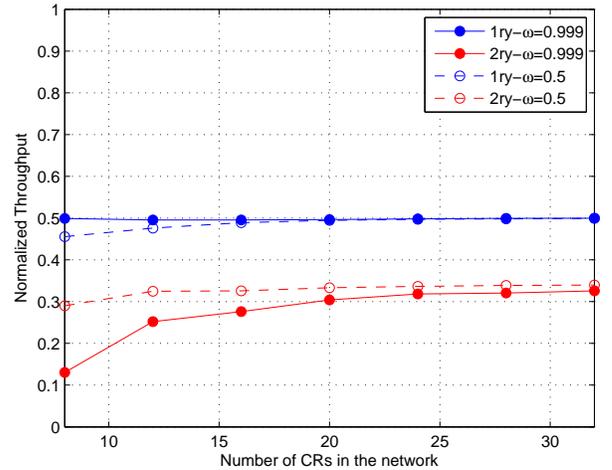}
	
\caption{Normalized primary and secondary throughputs vs. $M$, with $N=3$, $\sigma^{2}_{s_i}=2$, $K=8$, $\eta_p=1=\eta_s=1$, $R_p=R_s=1$, $\delta_p=\delta_s=0$ and $c=0.0001$ for $\omega=0.5$ and $\omega=0.999$. In the legend, ``1ry" refers to primary and ``2ry" refers to secondary. The scheme used to generate this figure is dynamic programming for throughout optimization.}
  \label{fig:throughput_w0.9_one_threshold}
\end{figure}

Figure \ref{fig:throughput_w0.9_one_threshold} shows the normalized primary and secondary throughputs for a weight $\omega=0.5$ and a weight $\omega=0.999$. The latter case is the case of interest in the context of cognitive radios given the privileges of the licensed primary users. The normalized throughput is obtained via simulating the system using the optimal thresholds. If the simulation is run over $Q$ time slots, the normalized secondary throughput is given by
\begin{equation}
\frac{1}{Q}\sum_{q=1}^{Q} I^{({\rm S})}_q R_{s} \left(1-\frac{\tau_N+k_{q}\tau}{\tau_s}\right)
\end{equation}
\noindent where $I^{({\rm S})}_q$ is equal to unity if the one of the secondary users transmits successfully over the $q^{th}$ time slot and zero otherwise, and $k_q$ is the number of probed sensors in the $q^{th}$ time slot. The normalized primary throughput is equal to $\frac{1}{Q}\sum_{q=1}^{Q} I^{({\rm P})}_q R_{p} $, where $I^{({\rm P})}_q$ is equal to unity when the primary user transmits successfully during the $q^{th}$ time slot and zero otherwise. It is clear from the figure that increasing the weight puts more emphasis on the primary throughput. The difference is, however, small in the case of a large number of sensors. For the rest of the results in this section and just for demonstration purposes, we use $\omega=0.5$.

\begin{figure}
	\centering
		\includegraphics[width=0.5\textwidth]{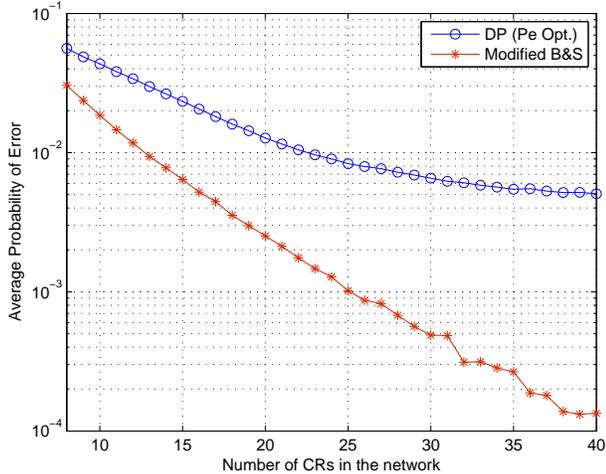}

		\caption{Average probability of error vs. $M$, $N=3$, $\sigma^{2}_{s_i}=2$, $K=8$, and $c=0.0001$.}
			\label{Pe_DP_BS}
\end{figure}

\begin{figure}
	\centering
	
		\includegraphics[width=0.5\textwidth]{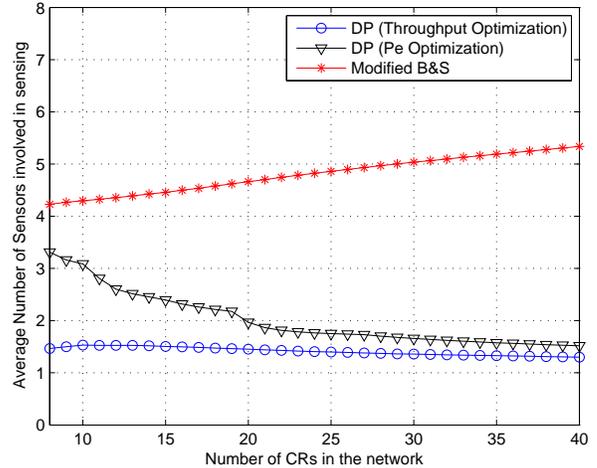}
		\caption{Average number of probed sensors for both DP schemes and the Modified B\&S scheme, vs. $M$, $N=3$, $\sigma^{2}_{s_i}=2$, $K=8$, and $c=0.0001$.}
  \label{Sensing_3schemes}
\end{figure}

\begin{figure}
	\centering
	
		\includegraphics[width=0.5\textwidth]{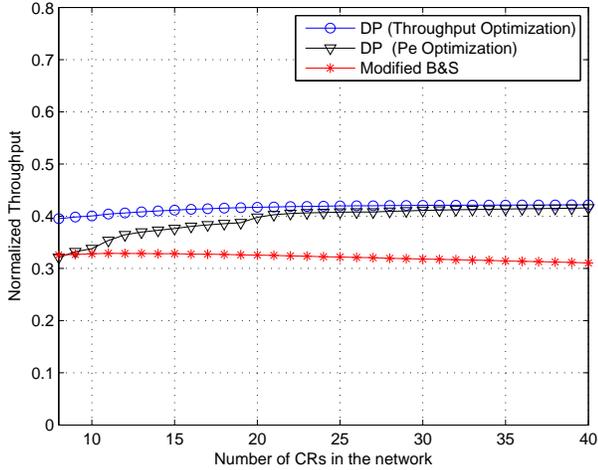}
		\caption{Normalized Weighted Sum Throughputs for both DP schemes and the Modified B\&S scheme, vs. $M$, $N=3$, $\sigma^{2}_{s_i}=2$, $K=8$, and $c=0.0001$.}
  \label{WS_3schemes}
\end{figure}

In Figure \ref{Pe_DP_BS}, it is clear that the modified scheme of \cite{sadler}, which we hereafter refer to as Modified B\&S scheme, achieves a lower probability of error than the dynamic programming scheme, denoted as DP scheme, since it achieves the same  probability of error as the MAP block detector. However, as illustrated in Figure \ref{Sensing_3schemes}, the number of sensors involved in the sensing process in the dynamic programming scheme is lower than that in the Modified B\&S scheme. The number of sensors actually converges slowly to unity using dynamic programming. Increasing the number of sensors in the network causes the ranked LLR observations to be more informative about the status of the channel. Therefore, when the DP scheme is used, the decision can be made after taking observations from one sensor on average, given that it has the maximum absolute LLR. However, for the Modified B\&S scheme, the sensing time increases as the number of CR's in the network increases. This follows from the fact that the greater the number of sensors in the network, the greater the distance between the two thresholds of comparison in (\ref{mod_sadler_thresholds}) causing more observations to be taken. Thus, the dynamic programming scheme represents a tradeoff of the sensing time with the error probability, via the $c$ parameter. The performance enhancement increases when the objective becomes to maximize the achievable weighted sum throughput as is evident from Figure \ref{Sensing_3schemes}. Figure \ref{WS_3schemes} demonstrates these results on the normalized secondary throughput, assuming unity primary and secondary transmission rates $R_p$ and $R_s$. For the normalized secondary throughput it is obvious from the simulation results that it converges to approximately $\pi_0 \left(1-\frac{\tau_N + \tau}{\tau_s}\right)$ in the DP scheme. This is the throughput for the hypothetical case of using just one ``perfect'' sensor that reveals that true state of the channel.

\begin{figure}[htbp]
	\centering
		\includegraphics[width=0.5\textwidth]{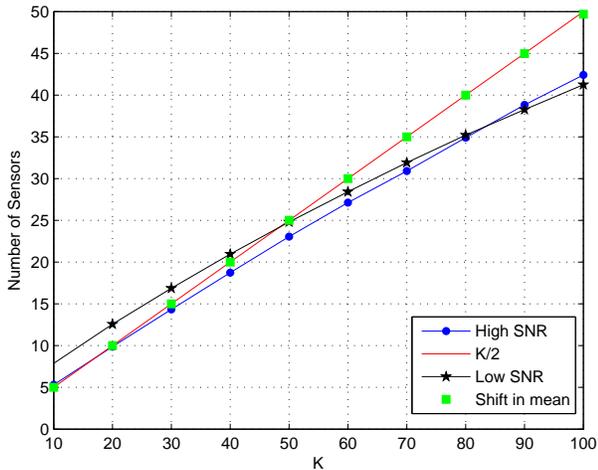}
		\caption{Number of sensors involved in sensing, for high SNR and low SNR versus different values for $K$, with $M=100$, $N=3$ and $\sigma^2_{s_i} =2$ for low SNR and $\sigma^2_{s_i} =50$ for high SNR. The ``shift in mean" case refers to using conditional distributions for local sensor measurements that, in contrast with (\ref{eq:xdef}), are Gaussian with the same variance and different means.}
	\label{fig:Sensing_vsK_low_highSNR}
\end{figure}

Using Modified B\&S scheme, it is noted that when the local sensors' SNR is high, the number of probed sensors is almost upperbounded by $\frac{K}{2}$ as shown in Figure \ref{fig:Sensing_vsK_low_highSNR}. For low SNR regime, the number of probed sensors is close to $\frac{K}{2}$, though may exceed it when the SNR is low depending on the value of $K$. If we use conditional distributions for the measurements at local sensors for which the correction term $\rho\left(y\right)$ in equation (\ref{mod_sadler_thresholds}) is zero, it can be shown analytically following a proof in \cite{sadler} that the maximum number of probed sensors is $\frac{K}{2}$. This is the case for example when the local measurements have conditional distributions that, in contrast with those in (\ref{eq:xdef}), are Gaussian with different means and the same variance. The simulation results for such ``shift in mean" case are also provided in Figure \ref{fig:Sensing_vsK_low_highSNR}.

\begin{figure}[htbp]
	\centering
		\includegraphics[width=0.50\textwidth]{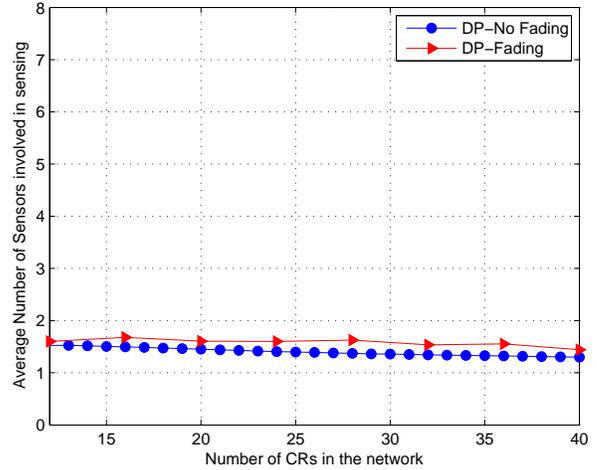}
		\caption{Average sensing time vs. $M$, with $N=3$, $\sigma^{2}_{s_i}=2$, $K=8$, using the DP scheme for error optimization under fading between the local sensors and the fusion center.}
	\label{fig:Sensing_fading}
\end{figure}

Taking into consideration the fading effect in the channels between the sensors and the fusion center as explained in Section \ref{sec:fading_channels}, only a subset of the total number of local detectors becomes involved in the sensing and reporting process according to the fading coefficients. This means that the fusion center takes longer time to detect the presence or absence of the primary user. Figure \ref{fig:Sensing_fading} shows the effect of fading, which is to slightly increase the average number of probed sensors. The parameters used to produce this figure are $b=20$, $W=50$ kHz, $\tau_{\rm b}=\frac{\tau}{2}=0.5$ msec, $\Gamma_i=2$, and $\frac{P_i}{\sigma_{\rm f}^{2}}=5$. Assuming that the channel gains between the sensors and the fusion center are i.i.d and are exponentially distributed with unity mean, we can easily find that $\delta$ is equal to $0.743$. Therefore, slightly less that three quarters of the number of sensors participate, on average, in the process of spectrum sensing. The observation that fading has a minor effect is just because of the simulation parameters. Note that in Figure \ref{Sensing_3schemes} the average number of sensors is already less than $2$ even for a relatively small $M$.

\subsection{One-Threshold Based Method}

As mentioned in Section \ref{sec:max_throughput}, instead of applying the two-threshold based scheme to the accumulated LLR observations at the fusion center, it can be sufficient to perform the comparison using one threshold only, when the sensing cost, $c$, is set to $0$. In fact, this requires consuming more mini-slots in sensing the channel, till a decision is made about the channel being idle. Otherwise, sensing continues till the end of the slot, declaring a busy channel.

\begin{figure}[tbp]
	\centering
		\includegraphics[width=0.50\textwidth]{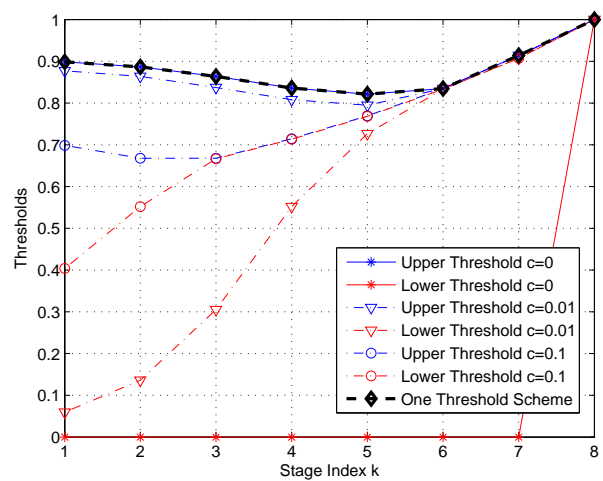}
	\caption{Thresholds of comparison for different costs vs. stage index $k$, with $K=8$, $M=10$, $\sigma^{2}_{s_i}=2$ and $N=3$.}
	\label{fig:One_vs_two_Thresholds}
\end{figure}

Figure \ref{fig:One_vs_two_Thresholds} shows the convergence of the two-threshold scheme to the one-threshold scheme at zero sensing cost, for the $M=10$ case. As $c$ decreases, the lower threshold is reduced at all stages. When $c=0$, the lower threshold is $0$ at all stages except for the last one at which a decision must be made. The fusion center therefore continues to take observations till the end of the slot if there is no sufficient statistics to declare $H_0$. At any stage up to $k=8$, the fusion center has only two choices; to declare $H_0$ or to probe more sensors, which is the main idea of the one-threshold scheme.

\begin{figure}[htbp]
	\centering
		\includegraphics[width=0.50\textwidth]{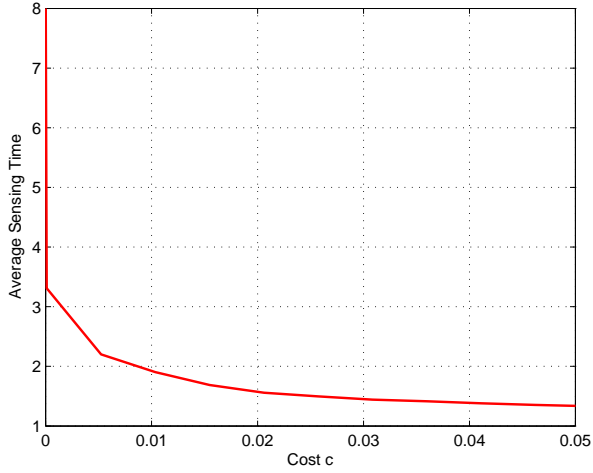}
	\caption{Average sensing time vs. $c$, with $N=3$, $\sigma^{2}_{s_i}=2$, $K=M=8$, using the DP scheme for error optimization.}
	\label{fig:Sensing_vs_C}
\end{figure}

To investigate the effect of the cost parameter $c$ on the sensing time, we consider the dynamic programming scheme with decision costs set for minimizing the average probability of error. Figure \ref{fig:Sensing_vs_C} shows that at $c=0$ the fusion center does not make a decision till the end of the slot. Increasing the cost gives the fusion center the chance to make a decision before the end of the slot allowing for secondary transmission. 

\section{Conclusion}
\label{sec:conc}

In this paper, we considered ordering the transmissions from cognitive users to the decision fusion center according to the information they carry about the status of the primary user. Restricting the transmissions to the highly informative set of observations, i.e., the LLR's with the highest magnitude, increases the time remaining in the time slot after making the decision, thereby boosting the throughput while protecting the primary signal. We devised a sequential scheme that achieves the optimum error probability of the MAP block detector, but makes the decision faster. We also devised a scheme based on dynamic programming that allows a trades-off between the average probability of error and the average sensing time. Simulation results have demonstrated that using dynamic programming approach to compute the thresholds required to make the decision at the fusion center yields a considerable enhancement in the performance in terms of reduced delay and increased secondary throughput.

Future work needs to address the situation of multiple primary channels. Also of interest is the design of the multiple access scheme that allows the sensor with the highest LLR in magnitude to seize the common control channel and transmit. Another front that needs further investigation is the channel access scheme and how the discovered spectrum opportunities are to be distributed among the secondary users. 

\appendices
\section{}
\label{sec:appendix2}

For the optimal probability of error MAP detector, using the best $K$ out of $M$ LLR magnitudes, the joint distribution of $K$ observations under hypothesis $H$ (either $H_0$ or $H_1$) is given by $f_{Y_{1}..Y_{K}\mid H}\left(y_1,y_2,..y_{K}\mid H\right)$. Since the sequence of observations forms a Markov chain (Theorem 2.4.3, \cite{Order_Statistics})
\begin{equation}
\begin{split}
f&_{Y_{1},Y_{2}...Y_{K}\mid H}(y_1,y_2,..y_{K}\mid H)= f_{Y^{[1]}}\left(Y^{[1]}=y_1\right) \cdot \\
& \prod_{j=2}^{K} f_{Y^{[j]}\mid Y^{[j-1]},H}(Y^{[j]}=y_j \mid Y^{[j-1]}=y_{j-1},H)
\end{split} 
\end{equation}

For symmetric sensors, the conditional distribution of the sensor of rank $j$ receiving an LLR value $Y^{[j]}=\alpha$ given that the sensor of rank $j-1$ has already received an LLR value $Y^{[j-1]}=\gamma$ is given by
\begin{equation}
\begin{split}
f&_{Y^{[j]}\mid Y^{[j-1]},H}(Y^{[j]}=\alpha \mid Y^{[j-1]}=\gamma,H)=\frac{(\beta_{\gamma,H})^{j-2}}{ \left(\beta_{\alpha,H}\right)^{j-2}} \cdot \\
& \frac{(1-\beta_{\alpha,H})^{M-j} M\left(M-1\right) \cdot f_{Y}(\alpha \mid H)f_{Y}(\gamma \mid H ){{M-2} \choose {j-2}}}{\left(1-\beta_{\alpha,H}\right)^{M-j+1} M \cdot f_{Y}\left(\gamma \mid H \right) {{M-1} \choose {j-2}}} = \\
& (M+1-j) f_{Y}(\alpha \mid H)\cdot \frac{\left(1-\beta_{\alpha,H}\right)^{M-j}}{\left(1-\beta_{\gamma,H}\right)^{M-j+1}}
\end{split}
\end{equation}
\noindent with $\beta_{x,H}$ defined as ${\rm Pr}\{|Y| \geq |x|\Big\vert H\}$, where random variable $Y$ is an LLR observation. Hence,
\begin{equation}
\begin{split}
f&_{Y_{1},Y_{2}...Y_{K}\mid H}(y_1,y_2,..y_{K}\mid H)=\\&
M f_Y\left(y_1 \big\vert H\right) \left(1-\beta_{y_1,H}\right)^{M-1}\cdot\\&
(M-1) f_{Y}(y_2 \big\vert H)\frac{\left(1-\beta_{y_2,H}\right)^{M-2}}{\left(1-\beta_{y_1,H}\right)^{M-1}} \cdot\\&
(M-2) f_{Y}(y_3 \big\vert H) \frac{\left(1-\beta_{y_3,H}\right)^{M-3}}{\left(1-\beta_{y_2,H}\right)^{M-2}} \cdot\\&
\cdot \cdot (M+1-K) f_{Y}(y_K \big\vert H) \frac{\left(1-\beta_{y_K,H}\right)^{M-K}}{\left(1-\beta_{y_{K-1},H}\right)^{M-K+1}}
\end{split}
\end{equation} 
Then, the likelihood ratio of the best $K$ out of $M$ LLR's for a block detector can be written as
\begin{equation}
\begin{split}
\label{prod_likelihoods}
& \frac{f_{Y_{1}..Y_{K}\mid H_1}(y_1,y_2,..y_{K}\mid H_1)}{f_{Y_{1}..Y_{K} \mid H_0}(y_1,y_2,..y_{K}\mid H_0)}=\\
& \frac{f_Y\left(y_1|H_1\right)..f_Y\left(y_K|H_1\right)\left(1-\beta_{y_{K},H_1}\right)^{M-K}}{f_Y\left(y_1|H_0\right)..f_Y\left(y_K|H_0\right)\left(1-\beta_{y_{K},H_0}\right)^{M-K}}
\end{split} 
\end{equation}
 
We now show that the sequential algorithm may operate on the received LLR's directly rather than computing their LLR's. This requires that the LLR is a strictly monotonic function of the observation. Consider observation $y$ with llr given by
\begin{equation}
\textrm{llr} = \log \frac{f_Y\left(y \mid H_1\right)}{{f_Y\left(y \mid H_0\right)}} = g(y)
\end{equation}
For a monotonic $g$ function, conditioned on $H$ (either $H_1$ or $H_0$),
\begin{equation}
f_{LLR}\left( \textrm{llr} \mid H\right) = f_{Y}\left(y \mid H\right) \Big\vert_{y=g^{-1}\left(\textrm{llr}\right)} \cdot \left|\frac{\textstyle{dg^{-1}\left(y\right)}}{\textstyle {d\textrm{llr}}}\right| 
\end{equation}

Then,
\begin{equation}  
\label{llr_of_llr}
\log \frac{f_{LLR}\left( \textrm{llr} \mid H_1\right)}{f_{LLR}\left( \textrm{llr} \mid H_0\right)}=
 \log \frac{f_{Y}\left(y \mid H_1\right) \Big\vert_{y=g^{-1}\left(\textrm{llr}\right)}}{f_{Y}\left(y \mid H_0\right) \Big\vert_{y=g^{-1}\left(\textrm{llr}\right)}} 
\end{equation}
But the right-hand-side of (\ref{llr_of_llr}) is given by
\begin{equation}
\begin{split}
\log \frac{f_{Y}\left(y \mid H_1\right) \Big\vert_{y=g^{-1}\left(\textrm{llr}\right)}}{f_{Y}\left(y \mid H_0\right) \Big\vert_{y=g^{-1}\left(\textrm{llr}\right)}} 
& = \log \frac{f_{Y}\left(g^{-1}\left(\textrm{llr}\right) \mid H_1\right)}{f_{Y}\left(g^{-1}\left(\textrm{llr}\right) \mid H_0\right)} \\
&= g\left(g^{-1}\left(\textrm{llr}\right) \right)  \\
&= \textrm{llr}
\end{split}
\end{equation}
Therefore,
\begin{equation}
\label{llr_proof}
\log \frac{f_{LLR}\left( \textrm{llr} \mid H_1\right)}{f_{LLR}\left( \textrm{llr} \mid H_0\right)}= \textrm{llr}
\end{equation}

If the LLR monotonicity assumption is not satisfied, the thresholds in (\ref{mod_sadler_thresholds}) can be modified as follows:
\begin{equation}
\begin{split}
&\hat{t}_k^{({\rm H})}=\log\frac{\pi_0}{1-\pi_0}-\min_{0 \leq y \leq |y_k|} \overline{\rho}\left(y\right)\\
&\hat{t}_k^{({\rm L})}=\log\frac{\pi_0}{1-\pi_0}-\max_{0 \leq y \leq |y_k|} \overline{\rho}\left(y\right)\\
\end{split}
\end{equation}
\noindent where 
\begin{equation}
\begin{split}
\overline{\rho}\left(y\right)=\left(M-K\right)\rho\left(y\right)+\left(K-k\right)\log\frac{f_Y\left(y\big\vert H_1\right)}{f_Y\left(y \big\vert H_0\right)}
\end{split}
\end{equation}
The proof that these thresholds achieve the same average probability of error if the best $K$ out of $M$ sensors are used follows the same outline as that presented in Section (\ref{sec:ord_scheme}). We focus here on the lower threshold. A decision in favor of $H_0$ is made if the metric $\sum_{m=1}^{k}\log\frac{f_Y\left(y_m|H_1\right)}{f_Y\left(y_m|H_0\right)}$ gets below $\hat{t}_k^{({\rm L})}$. If $\sum_{m=1}^{k}\log\frac{f_Y\left(y_m|H_1\right)}{f_Y\left(y_m|H_0\right)} < \hat{t}_k^{({\rm L})}$,
\begin{equation}
\begin{split}
&\sum_{m=1}^{k}\log\frac{f_Y\left(y_m|H_1\right)}{f_Y\left(y_m|H_0\right)}<\log\frac{\pi_0}{1-\pi_0}-\max_{0 \leq y \leq |y_k|} \overline{\rho}\left(y\right)<\\
& \log\frac{\pi_0}{1-\pi_0}-\sum_{m=k+1}^{K}\mbox{ \hspace{-5 mm} log}\frac{f_Y\left(y_m|H_1\right)}{f_Y\left(y_m|H_0\right)}-\left(M-K\right)\max_{0 \leq y \leq |y_k|} {\rho}\left(y\right)
\nonumber
\end{split}
\end{equation}
\noindent Then
\begin{equation}
\begin{split}
\sum_{m=1}^{K}\mbox{\hspace{-2.2 mm} log}\frac{f_Y\left(y_m|H_1\right)}{f_Y\left(y_m|H_0\right)}&<\log\frac{\pi_0}{1-\pi_0}-\left(M-K\right)\max_{0 \leq y \leq |y_k|} {\rho}\left(y\right)\\
&<\log\frac{\pi_0}{1-\pi_0}-\left(M-K\right)\max_{0 \leq y \leq |y_K|} {\rho}\left(y\right)\\
&<\log\frac{\pi_0}{1-\pi_0}-\left(M-K\right)\log\frac{1-\beta_{y_{K},H_1}}{1-\beta_{y_{K},H_0}}
\end{split}
\end{equation}
\noindent which is the MAP rule for a decision based on a block of the $K$ most reliable measurements among $M$ observations.

\section{}
\label{sec:appendix3}
In this appendix we prove that when the throughput terms are preserved while all other costs, including parameter $c$, are set to zero, the lower detection threshold for the sequential procedure becomes always zero except for the last stage. This means that, from a throughput point of view, the optimal decisions based on the observations are either to declare $H_0$ and then select a cognitive terminal for channel access, or to continue taking more samples so long as less than $K$ LLR values have been gathered. 

We start by proving the concavity of function $J^{K}_{k}\left(\pi_{k}\right)$ using induction. At the last stage
\begin{eqnarray}
J^{\left(K\right)}_{K}(\pi_K) &=& \text{min}\; \bigg\{ \:-r_s\left(1-\frac{\tau_N+K\tau}{\tau_s}\right)\;\pi_K, \nonumber\\
& & \;\;\;\;\;\;\;\; -r_p\;\left(1-\pi_K\right) \bigg\}
\label{qwe3}
\end{eqnarray}
\noindent where $r_s=\left(1-\omega\right)R_s$ and $r_p=\omega R_p$. For $k<K$,
\begin{eqnarray}
J^{\left(K\right)}_{k}(\pi_k) &=& \text{min}\; \bigg\{ \:-r_s\left(1-\frac{\tau_N+k\tau}{\tau_s}\right)\;\pi_k, \nonumber\\
& & \;\;\;\;\;\;\;\; -r_p\;\left(1-\pi_k\right),\;c+\Psi_k\left(\pi_k\right) \bigg\}
\label{qwe2}
\end{eqnarray}
\noindent where $\Psi_k\left(\pi_k\right)$ is given by
\begin{equation}
\begin{split}
\Psi_k\left(\pi_k\right)=&\int J^{K}_{k+1}\left( \frac{\pi_{k}f_0}{\pi_{k}f_0+{(1-\pi_{k})f_1}}\right)\cdot \\
& \Big[\pi_{k}f_0+\left(1-\pi_k\right)f_1\Big]dY^{[k+1]}
\end{split}
\end{equation}
\noindent Note that we use $f_0$ in place of $f_{Y^{[k+1]}}\left(y \mid H_0\right)$ and $f_1$ in place of $f_{Y^{[k+1]}}(y \mid H_1)$ to simplify notation. Parameter $c$ is kept in expression (\ref{qwe2}) and is set to zero after the concavity of $J^{K}_{k}\left(\pi_{k}\right)$ is established.

If $\Psi_k\left(\pi_k\right)$ is concave, then $c+\Psi_k\left(\pi_k\right)$ is concave and, consequently, $J^{K}_{k}\left(\pi_{k}\right)$ is concave because $J^{K}_{k}\left(\pi_{k}\right)$ is the minimum of three concave terms \cite{boydconvex}. We now prove the concavity of $\Psi_k\left(\pi_k\right)$ under the assumption that $J^{K}_{k+1}\left(\pi_{k+1}\right)$ is concave. Function $\Psi_k\left(\pi_k\right)$ is concave if it satisfies the following inequality
\begin{equation}
\label{inequality_concave}
\kappa \Psi_k\left(x\right) + \left(1-\kappa\right)\Psi_k\left(z\right) \leq \Psi_k\left(\kappa x +\left(1-\kappa\right)z\right)
\end{equation} 
\noindent $\forall \kappa \in \left[0,1\right]$ \cite{boydconvex}. Let
\begin{equation}
\epsilon=\kappa \left[x f_0+\left(1-x\right)f_1\right]+\left(1-\kappa\right) \left[z f_0+\left(1-z\right)f_1\right]
\end{equation}
The right-hand-side of (\ref{inequality_concave}) then can be written as
\begin{equation}
\begin{split}
&\int \biggl\{ \frac{\kappa}{\epsilon} \Big[x f_0+\left(1-x\right)f_1\Big] J^{K}_{k+1}\left( \frac{x f_0}{x f_0+{(1-x)f_1}}\right)+ \\
&\frac{1-\kappa}{\epsilon} \Big[z f_0+\left(1-z\right)f_1\Big] J^{K}_{k+1}\left( \frac{z f_0}{z f_0+{(1-z)f_1}}\right) \biggr\} \epsilon dY^{[k+1]}
\end{split} 
\end{equation}
\noindent Noting that $\frac{\kappa}{\epsilon} \left[x f_0+\left(1-x\right)f_1\right]+\frac{1-\kappa}{\epsilon} \left[z f_0+\left(1-z\right)f_1\right]=1$ and using the assumption that $J^{K}_{k+1}$ is concave, we obtain
\begin{equation}
\label{inequality_a}
\begin{split}
&\kappa \Psi_k\left(x\right) + \left(1-\kappa\right)\Psi_k\left(z\right) \leq \\
& \int J^K_{k+1}\left(\frac{\kappa}{\epsilon}x f_0+\frac{1-\kappa}{\epsilon}z f_0\right) \epsilon dY^{[k+1]}
\end{split}
\end{equation}
\noindent Factor $\epsilon$ can be expressed as $\epsilon=\bar{\epsilon} f_0 + \left(1-\bar{\epsilon}\right) f_1$, where $\bar{\epsilon}=\kappa x +\left(1-\kappa\right)z$. Using this, inequality (\ref{inequality_a}) becomes
\begin{equation}
\label{inequality_b}
\begin{split}
&\kappa \Psi_k\left(x\right) + \left(1-\kappa\right)\Psi_k\left(z\right) \leq \\
& \int J^K_{k+1}\left(\frac{\bar{\epsilon}f_0}{\bar{\epsilon}f_0+\left(1-\bar{\epsilon}\right)f_1}\right) \Big[\bar{\epsilon} f_0 + \left(1-\bar{\epsilon}\right) f_1\Big] dY^{[k+1]} \\
& = \Psi_k\left(\bar{\epsilon}\right)=\Psi_k\left(\kappa x +\left(1-\kappa\right)z\right)
\end{split}
\end{equation}
\noindent Therefore, $\Psi_k\left(\pi_k\right)$ is concave in $\pi_k$ assuming that $J^K_{k+1}\left(\pi_{k+1}\right)$ is concave. It is evident from (\ref{qwe3}) that $J^{K}_K\left(\pi_K\right)$ is concave since it is the minimum of two affine terms. By induction both $\Psi_k\left(\pi_k\right)$ and $J^{K}_k\left(\pi_k\right)$ are concave in $\pi_k$ for $k<K$.  

We now study the value of $J^K_{k}\left(\pi_k\right)$ at $\pi_k=0$. It is clear that $J^{K}_K\left(0\right)=-r_p$. Since $\Psi_{K-1}\left(0\right)=\int J^{K}_{K}\left(0\right) f_1 dY^{[K]}=-r_p$, then $J^K_{K-1}\left(0\right)=\min\{0,-r_p,c-r_p\}=-r_p$ for $c \geq 0$. It is straightforward to show that $\Psi_k\left(0\right)=J^K_{k}\left(0\right)=-r_p$ for $1 \leq k < K$. 

When $\pi_k=0$, the minimum of the three terms of (\ref{qwe2}) is $-r_p$ obtained from the second term $-r_p\left(1-\pi_k\right)$ for $c>0$. The value of the derivative of $J^K_{k}\left(\pi_k\right)$ at $\pi_k=0$ is equal to the derivative of the term $-r_p\left(1-\pi_k\right)$ at $\pi_k=0$, which is equal to $r_p$. The derivative of $\Psi_k\left(\pi_k\right)$ is given by
\begin{equation}
\label{epsi_slope}
\begin{split}
\frac{d\Psi_k}{d\pi_k}=&\int \frac{dJ^K_{k+1}}{d\pi_{k+1}} \frac{d\pi_{k+1}}{d\pi_k} \Big[\pi_k f_0 + \left(1-\pi_k\right)f_1\Big] dY^{[k+1]} \\
& + J^K_{k+1}\left(\pi_{k+1}\right)\Big[f_0 - f_1\Big] dY^{[k+1]} 
\end{split}
\end{equation}
\noindent since $\pi_{k+1}=\frac{\pi_k f_0}{\pi_k f_0 + \left(1-\pi_k \right)f_1}$, $\pi_{k+1}=0$ when $\pi_k=0$ and it is straightforward to show the $\frac{d\pi_{k+1}}{d\pi_k}$ at $\pi_k=0$ is equal to $\frac{f_0}{f_1}$. Hence,
\begin{equation}
\left. \frac{d\Psi_k}{d\pi_k}\right\vert_{\pi_k=0}=\int \left(r_p \frac{f_0}{f_1} f_1 -r_p \left[f_0 - f_1\right]\right)dY^{[k+1]}=r_p
\end{equation}
\noindent Since $\Psi_k$ is concave, it satisfies the following inequality \cite{boydconvex}
\begin{equation}
\Psi_k\left(z\right) \leq \Psi_k\left(x\right)+\left. \frac{d\Psi_k}{dx}\right\vert_{x}\left(z-x\right)
\end{equation}
\noindent For $x=0$ and $z=\pi_k$, $\Psi_k\left(\pi_k\right) \leq -r_p +r_p \pi_k =-r_p\left(1-\pi_k\right)$. This means that when $c=0$, the third term in (\ref{qwe2}) is always less than or equal the second term $-r_p\left(1-\pi_k\right)$. Hence, the cost of getting one more sample is always less than or equal to the cost of declaring $H_1$. In other words, when $c=0$, the lower decision threshold is zero at all the stages with the exception of the last one because it is less costly to continue sampling rather than declaring $H_1$.

\bibliographystyle{IEEEbib}
\bibliography{MyLib}

\end{document}